\documentclass[aps,prl,twocolumn,superscriptaddress,preprintnumbers,amsmath,amssymb,longbibliography,floatfix,amsfonts,lengthcheck]{revtex4-1}

\usepackage{ulem}  
\usepackage{graphicx}
\usepackage{mathtools}
\usepackage{xcolor}
\usepackage{multirow}
\usepackage{array}
\usepackage{textgreek}
\usepackage{bm}
\usepackage{contour}
\usepackage{braket}

\renewcommand{\vec}[1]{\bm{#1}}
\newcommand{\tens}[1]{\mbox{\textsf{\textbf{#1}}}}
\newcommand{\At}{\mathrm{A}}
\newcommand{\Dt}{\mathrm{D}}

\newcommand{\ev}{\vec{e}}

\newcommand{\ee}{\mathrm{e}}

\newcommand{\om}{\omega}

\newcommand{\im}{\mathrm{i}}

\newcommand{\sprod}{\!\cdot\!}
\newcommand{\vprod}{\!\times\!}

\begin{document}

\title{Photoelectron circular dichroism of a chiral molecule induced by resonant interatomic Coulombic decay  from an antenna atom} 

\author{Stefan Yoshi Buhmann}
\email{stefan.buhmann@uni-kassel.de}\affiliation{\kassel}
\author{Andreas Hans}
\email{hans@physik.uni-kassel.de}\affiliation{\kassel}
\author{Janine C. Franz}
\affiliation{\kassel}
\affiliation{\bordeaux}
\author{Philipp V. Demekhin}
\email{demekhin@physik.uni-kassel.de}\affiliation{\kassel}

\newcommand{\kassel}{Institut f\"{u}r Physik und CINSaT, Universit\"{a}t Kassel, Heinrich-Plett-Stra{\ss}e 40, 34132 Kassel, Germany}
\newcommand{\bordeaux}{Universit\'e de Bordeaux, CNRS, LOMA, UMR 5798, F-33405 Talence, France}

\date{\today}

\begin{abstract}
    We show that a nonchiral atom can act as an antenna to induce a photoelectron circular dichroism in a nearby chiral molecule in a three-step process: The donor atom  (antenna)  is initially resonantly excited by circularly polarized radiation. It then transfers its excess energy to the acceptor molecule by means of resonant interatomic Coulombic decay. The latter finally absorbs the energy and emits an electron which exhibits the aforementioned circular dichroism in its angular distribution. We study the process on the basis of the retarded dipole--dipole interaction and report an asymptotic analytic expression for the distance-dependent chiral asymmetry of the photoelectron as induced by resonant interatomic Coulombic decay for random line-of-sight and acceptor orientations. In the nonretarded limit, the predicted chiral asymmetry is reversed as compared to that of a direct photoelectron circular dichroism of the molecule.   
\end{abstract}

\maketitle


\paragraph*{Introduction.}\label{sec:Introduction}
Chirality describes a symmetry property in which a mirror image of an object cannot be brought into overlap with the original by means of rotation and translation. Chiral molecules play a central role in the biosphere. Most of the larger biomolecules are chiral, and, intriguingly, their two isomers of opposite handedness (called enantiomers) usually differ in their biological effect and function, a phenomenon known as the `homochirality of life' \cite{Meierhenrich2008}. Photoelectron circular dichroism (PECD) is a frequently used tool for the recognition of chiral molecules in the gas phase. It describes the enantio- and helicity-dependent forward-backward asymmetric angular distribution of photoelectrons emitted by randomly oriented chiral molecules with respect to the direction of propagation of the ionizing light \cite{Ritchie1976}. PECD was first observed experimentally in  one-photon ionization with synchrotron radiation \cite{Bowering2001,Garcia2003,Hergenhahn2004}. Nowadays, it is known to be a universal phenomenon occurring in different photoionization regimes \cite{Powis2008a,Nahon2015,Beaulieu2016,Wollenhaupt2016}. All observations have in common that the effect is observable for photoelectrons with a relatively low kinetic energy of about $0-20\,$eV. The investigation of PECD of isolated chiral molecules is essential for a fundamental understanding of chirality. However, in reality, chiral molecules do not appear in the gas phase but are rather embedded in an environment. For instance, PECD in clusters of chiral molecules was addressed in pioneering works ~\cite{Powis2014,Hartweg2021}. Very recently, PECD was observed in achiral chromophores induced by bound chiral molecules \cite{Rouquet2023}. However, such studies of PECD in realistic and extended environments are still very rare.

An important aspect of the photophysics of extended systems like clusters and liquids, which was discovered and extensively explored during the last decades, is that in addition to conventional (direct) photoemission, other non-local processes contribute to the observed electron spectra. In particular, a phenomenon called interatomic (intermolecular) Coulombic decay (ICD) in a variety of realizations has attracted significant attention. ICD is a non-local autoionization mechanism, in which electronic excess energy of a quantum system is transferred to a remote neighbor, and it is used to ionize the latter \cite{Cederbaum1997}. It has been found to be omnipresent in dense media, from prototypical rare-gas clusters to aqueous solutions \cite{Jahnke2020}. An important aspect of ICD is that it decisively influences the photochemistry of an excited system through the ionization of a neighbor. The created final electronic states of the whole system and respective charge distributions determine the resulting nuclear dynamics. Often both the energy donor and acceptor become charged after ICD, leading to rapid fragmentation via Coulomb explosion \cite{Jahnke2020,Jahnke2004} and, thus, to a disintegration of a system. In addition, an ICD-like energy transfer can even lead to neutral dissociation \cite{Cederbaum2020}. After the discovery of the ICD as a relaxation pathway for photoionized systems, it became clear that this non-local mechanism can straightforwardly be transferred to resonant excitations  \cite{Barth2005,Gokhberg2006,Knie2014}. Intriguingly, through  resonant interatomic Coulombic decay (rICD) , the ionization cross section of a quantum system can be efficiently enhanced if tuning the photon energy to be resonant with the light absorption by a neighboring `antenna' \cite{Najjari2010,Trinter2013,Hans2019}.

So far, PECD as an evidence for the system's chirality on the one hand and non-local ICD-like energy-transfer phenomena on the other have been considered as separate physical phenomena described by different theoretical frameworks: The former relies on a variety of many-electron intramolecular physics and quantum chemistry approaches, while the latter can alternatively be studied via intermolecular quantum chemistry \cite{Santra02} or with macroscopic quantum electrodynamics (QED) ---a theory of the interaction of molecular systems with photons that includes effects of retardation and dielectric environments \cite{Hemmerich2018}. In particular, this means that the benefits of PECD as a sensitive purely electric diagnostic tool for the discrimination of chiral molecules have remained inaccessible to the field of resonance energy transfer, where only relatively weak discriminatory processes induced by optical activity have been considered so far \cite{Craig98b,Franz24}. In this theoretical work, we combine macroscopic QED and intramolecular physics to show, that a non-local resonant energy transfer is enantio-selective and can give rise to a PECD-like effect. In particular, we use the example of rICD following absorption of circularly polarized light by an antenna atom and demonstrate that a sizable PECD-like asymmetry can be observed in the respective rICD-electron spectra. The predicted phenomenon substantially enriches both fields and, depending on the point of view or discipline, can  be viewed as a `chiral energy transfer  without optical activity' or, alternatively,  as an `antenna-induced PECD'.


 \begin{figure}
    \includegraphics[width=1.\columnwidth]{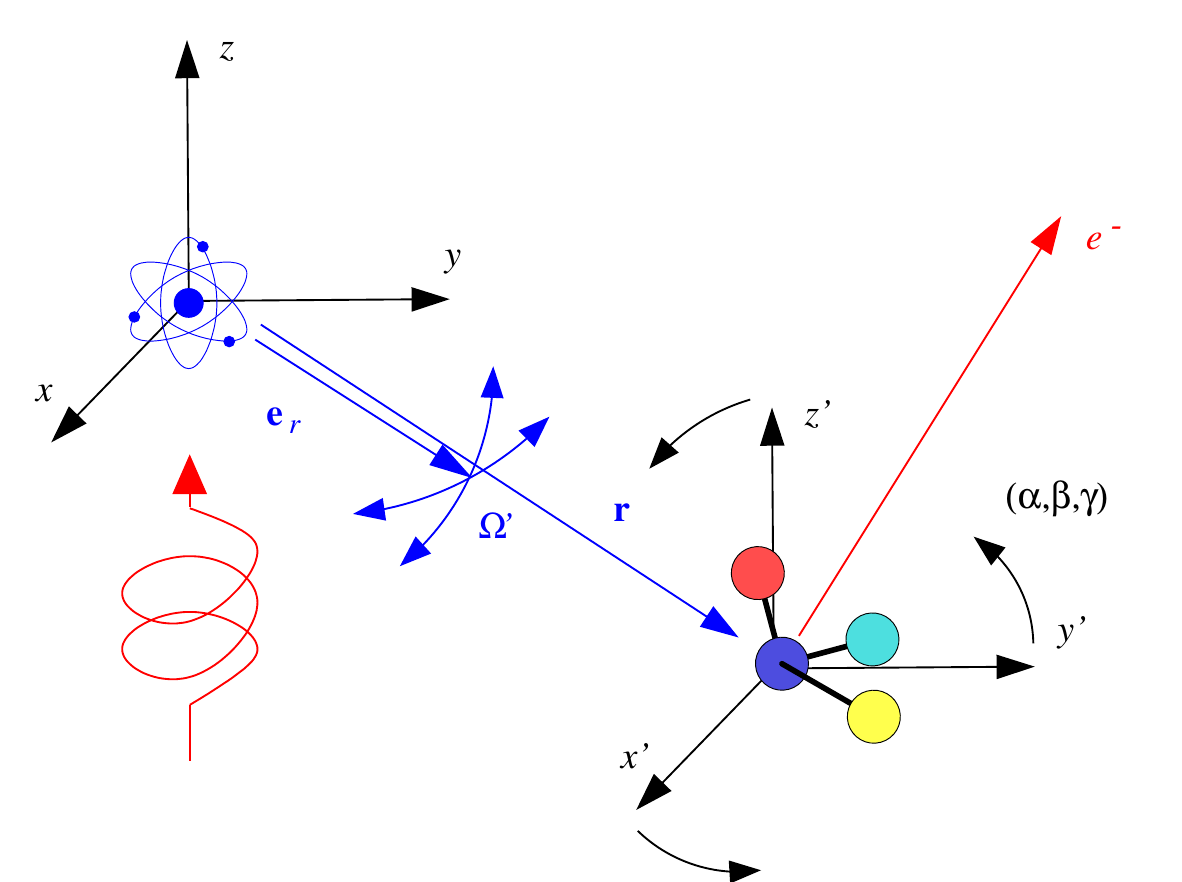}
    \caption{Sketch of the antenna-induced PECD. An atom is excited by circularly polarized light. It transfers its excess energy to a chiral molecule at a  relative position $\vec{r}$. The latter emits a photoelectron whose angular distribution exhibits circular dichroism. We average over the line-of-sight orientations between the constituents ($\Omega^\prime$) and the relative molecular orientations $(\alpha,\beta,\gamma)$.}
    \label{fig:Configuration}
    \end{figure}


\paragraph*{Scenario.}\label{sec:Scenario}
The scenario we are considering theoretically is depicted in Fig.~\ref{fig:Configuration}, where the process of antenna-induced PECD takes place in three steps:
\begin{itemize}
    \item[(i)] An achiral donor atom has been resonantly excited from its ground state $\ket{\alpha}$ to an excited electronic state $\ket{\gamma}$ by absorption of circularly polarized light. By selection rules for atomic transitions, the light's polarization $q=\pm1$ creates a unique single state $\ket{\gamma}$ of the donor (see example below). The excitation frequency and electric-dipole matrix element of the donor  are given by $\omega_\Dt = (E_{\gamma}^\Dt - E_{\alpha}^{\Dt})/\hbar > 0$ and $\vec{d}_{\gamma\alpha}^\Dt\equiv\langle\gamma|\hat{\vec{d}}^\Dt|\alpha\rangle =   -d^\Dt_{\pm 1} \ev_{\mp 1}$, respectively. The latter has been expressed via its spherical vector components $d^\Dt_{\pm 1}$ (see End Matter), which depend on the helicity of the exciting light. Assuming the ground state to exhibit an orbital angular momentum $\ell=0$ and  projection $m=0$ onto the laboratory-frame $z$-axis (defined by the direction of the propagation of light), the excited state will exhibit quantum numbers $\ell=1$ and  $m=\pm 1$. 
    \item[(ii)] For simplicity, we assume that the donor atom can undergo a single downwards transition \mbox{$\gamma\to\alpha$}, characterized by the same frequency $\omega_\Dt$, and transfers its energy to a nearby chiral acceptor molecule at a relative position $\vec{r}$ by means of rICD. The electric-dipole transition matrix element is given by $\vec{d}_{\alpha \gamma}^\Dt\equiv\langle\alpha|\hat{\vec{d}}^\Dt|\gamma\rangle =  \langle\gamma|\hat{\vec{d}}^\Dt|\alpha\rangle^\ast= d^{\Dt*}_{\pm 1} \ev_{\pm 1}$.   
    \item[(iii)] The acceptor molecule, initially in its ground electronic state, absorbs the transferred energy, whereby an electron undergoes a transition from a bound state $\ket{\delta}=\ket{\phi_0}$ to a continuum state $\ket{\psi^-_{\vec{k}}}$ with a wave vector $\vec{k}$, so that the final state of the molecule is given by $\ket{\beta}=\ket{\psi_i\psi^-_{\vec{k}}}$ with $\ket{\psi_i}$ denoting the state of the remaining molecular ion. This photoionization transition is characterized by a frequency $\omega_\At = (E_{\beta}^\At - E_{\delta}^{\At})/\hbar > 0$ and an electric-dipole matrix element $\vec{d}^\At_{\beta\delta}=\bra{\psi_i\psi^-_{\vec{k}}}\hat{\vec{d}}^\At\ket{\phi_0}$. 
\end{itemize}
For generality, we assume a random intermolecular orientation of the donor and acceptor, as given by both, the  line-of-sight orientation between the constituents ($\Omega^\prime$) and the relative molecular orientation angles $(\alpha,\beta,\gamma)$. Because the relative orientation between them is not fixed, the considered scenario is very different from a conventional resonant Auger decay \cite{Hartmann2019} where donor and acceptor form a bound chiral system. Although direct photoionization of a chiral molecule by the incoming photon takes place as well, we will focus on the described above rICD  mechanism, which is by far dominant at resonant photon energies (e.g., by a factor of about 60 in a previously studied case \cite{Trinter2013}).

    
\paragraph*{Resonant interatomic Coulombic decay.}
A general expression for the rate of resonance energy transfer processes of the considered kind for isotropic molecules was derived in Ref.~\cite{Hemmerich2018} from the multipolar molecule--field coupling Hamiltonian of macroscopic QED in electric-dipole approximation using second-order perturbation theory. In the End Matter section, we generalize this calculation to the anisotropic case required here. The result  
\begin{equation}
\label{eq:1}
 \Gamma_{{\gamma\delta}}= \frac{2\pi{\omega_\Dt^4}}{{\hbar\varepsilon_0^2c^4}}\sum_{\beta} {\vert \vec{d}_{\beta \delta}^\At\sprod \tens{G} (\vec{r}_\At,\vec{r}_\Dt,\omega_\Dt) \sprod\vec{d}_{\alpha \gamma}^\Dt \vert}^2 \delta({\hbar}\omega_\Dt\!-\!{\hbar}\omega_\At)
\end{equation}
is valid for relatively large distances such that electronic wave-function overlap between donor and acceptor can be neglected. It hence neglects exchange interactions and charge-transfer processes, but accounts for both retardation and the possible presence of magneto-electric environments via the Green tensor of the electromagnetic field. In the absence of such environments, the Green tensor is given by its known free-space form \cite{Hemmerich2018}
\begin{equation}
\label{eq:2}
    \tens{G}(\vec{r}_\At,\vec{r}_\Dt,\omega_\Dt) = - \frac{c^2 \ee^{\im \om_\Dt r/c}}{4 \pi \om^2_\Dt r^3} \left( f \tens{I} - g \ev_r \otimes \ev_r \right),
\end{equation}
with $\vec{r}= \vec{r}_\At - \vec{r}_\Dt{\neq\mathbf{0}}$, $r=|\vec{r}|$, $\ev_r=\vec{r}/r$, $\tens{I}$ denoting the unit tensor, and
\begin{equation}
\label{eq:3fg}
\begin{split}
    f &= f\left(\frac{\omega_{\Dt}}{c} \right) = 1-\im \frac{\omega_{\Dt}}{c} - \left( \frac{\omega_{\Dt}}{c} \right)^2, \\
    g &= g\left(\frac{\omega_{\Dt}}{c}\right) = 3-3\im \frac{\omega_{\Dt}}{c} - \left( \frac{\omega_{\Dt}}{c} \right)^2.
\end{split}
\end{equation}
When the donor--acceptor distance is much smaller than the wavelength associated with the electronic transitions such that $\omega_\Dt r/c \ll 1$,  the following approximations can be applied to the Green tensor: $f \approx f(0) = 1$, $g \approx g(0) = 3$, and $   \ee^{\im \omega_\Dt r/c } \approx 1$. In this nonretarded limit, the energy transfer rate reduces to the F\"{o}rster rate \cite{Forster1948} and is essentially given by the square of the matrix element of the electrostatic dipole--dipole interaction
\begin{equation}
\label{eq:9}
    \hat{V}=\frac{\hat{\vec{d}}^\At\cdot\hat{\vec{d}}^\Dt-3\bigl(\hat{\vec{d}}^\At\cdot\vec{e}_r\bigr)\bigl(\hat{\vec{d}}^\Dt\cdot\vec{e}_r\bigr)}{4\pi\varepsilon_0r^3} 
\end{equation}
through Fermi's golden rule
\begin{equation}
   \Gamma_{{\gamma\delta}}=\frac{2\pi}{\hbar}\sum_{\beta}|\bra{\beta}\bra{\alpha}\hat{V}\ket{\delta}\ket{\gamma}|^2\delta({\hbar}\omega_\Dt-{\hbar}\omega_\At). \label{eq:FGR}
\end{equation}
After summing over all final free-electron states of the acceptor, the above energy-transfer rate yields the total transition rate of the rICD process.

Below, we consider partial rates for given electron emission angles  $\Omega = (\theta,\varphi)$ in the laboratory frame.  Therefore, by adapting the general formula~(\ref{eq:1}) to our scenario, a summation over the photoelectron emission directions $\Omega$ in the final acceptor states $\ket{\beta}$ must be omitted. Resolving the $\delta$-function according to $\delta(\omega_\Dt - \omega_\At)\Rightarrow \varepsilon= \hbar\omega_\Dt - \mathrm{IP}$ with $\varepsilon=k^2/2m$ being kinetic energy of the emitted electron, and $\mathrm{IP}$ denoting the ionization potential, we arrive at an angle-resolved rate
\begin{equation}
    \Gamma_\pm(\Omega)= {\frac{2\pi\omega_\Dt^4}{\hbar\varepsilon_0^2c^4}} \sum_{\psi_i} {\vert  \vec{d}_{\beta \delta}^\At\sprod \tens{G} (\vec{r}_\At,\vec{r}_\Dt,\omega_\Dt) \sprod \vec{d}_{\alpha \gamma}^\Dt \vert}^2.
\end{equation}
Here, the subscript index $\gamma$ of the rate from Eq.~(\ref{eq:1}) is replaced  by the index $\pm$, which explicitly indicates helicity of the exciting light, and the subscript $\delta$ indicating the ground state of the acceptor molecule is omitted for brevity. After using the free-space Green tensor~(\ref{eq:2}), we obtain 
\begin{equation}
\begin{split}
    \Gamma_\pm(\Omega)=& {\frac{1}{8\pi \varepsilon_0^2\hbar r^6}}\sum_{\psi_i}\bigl(\vert f \vert^2  \vec{d}_{\beta \delta}^\At \cdot\vec{d}_{\alpha \gamma}^\Dt \vec{d}_{\beta \delta}^{\At*}\cdot\vec{d}_{\alpha \gamma}^{\Dt*}  \\
    &-gf^* \vec{d}_{\beta \delta}^\At\cdot {\ev_r  \ev_r} \cdot \vec{d}_{\alpha \gamma}^\Dt \vec{d}_{\beta \delta}^{\At*}\cdot \vec{d}_{\alpha \gamma}^{\Dt*} \\
    &-fg^*\vec{d}_{\beta \delta}^\At \cdot \vec{d}_{\alpha \gamma}^\Dt \vec{d}_{\beta \delta}^{\At*}\cdot {\ev_r \ev_r} \cdot \vec{d}_{\alpha \gamma}^{\Dt*} \\
    & +\vert g \vert^2  \vec{d}_{\beta \delta}^\At\cdot {\ev_r  \ev_r} \cdot \vec{d}_{\alpha \gamma}^\Dt \vec{d}_{\beta \delta}^{\At*}\cdot {\ev_r \ev_r} \cdot \vec{d}_{\alpha \gamma}^{\Dt*}\bigr).
\end{split} \label{eq:strt}
\end{equation}
Equation~(\ref{eq:strt}) is our starting point to compute the helicity-dependent angle-resolved transition rate for the considered three-step process (i--iii) and the antenna-induced PECD.


\paragraph*{Average over the intermolecular line-of-sight orientations.}
So far, the rate (\ref{eq:strt})  explicitly depends on the orientation of the intermolecular line-of-sight $\vec{e}_r$ and that of the acceptor molecule $(\alpha,\beta,\gamma)$ with respect to the laboratory frame, as defined by the wave vector of the exciting light (see Fig.~\ref{fig:Configuration}). In the derivations below, we consider these two orientations to be independent of each other. This is justified, unless donor and acceptor form a tightly bound system, where the donor atom is fixed at a particular site of the acceptor molecule. We, thus, first average over the orientation of the line-of-sight angles ($\Omega^\prime$) by using the identities \cite{Craig1998}: \mbox{$\overline{\ev_r \otimes \ev_r} \equiv (1/4\pi) \int \text{d}\Omega'\, \ev_r \otimes \ev_r = \frac{1}{3}\tens{I}$} and
\mbox{$\overline{(\ev_r \otimes \ev_r \otimes \ev_r \otimes \ev_r)}_{ijkl} 
 = \frac{1}{15}\left(\delta_{ij}\delta_{kl} + \delta_{ik}\delta_{jl} + \delta_{il}\delta_{jk} \right)$}. This leads to
\begin{multline}
\label{eq:11}
    \Gamma_\pm(\Omega)= {\frac{1}{8\pi \varepsilon_0^2\hbar r^6}}\sum_{\psi_i}\bigl[\bigl( \vert f\vert^2 -\tfrac{1}{3}gf^* - \tfrac{1}{3}fg^* + \tfrac{1}{15}\vert g \vert^2 \bigr) \\
    \times\vec{d}_{\beta \delta}^\At \cdot\vec{d}_{\alpha \gamma}^\Dt  \vec{d}_{\beta \delta}^{\At*} \cdot \vec{d}_{\alpha \gamma}^{\Dt*}  
    +\tfrac{1}{15} \vert g \vert^2 \vec{d}_{\beta \delta}^\At  \cdot \vec{d}_{\beta \delta}^{\At*}\vec{d}_{\alpha \gamma}^\Dt \cdot \vec{d}_{\alpha \gamma}^{\Dt*}  \\
    +\tfrac{1}{15} \vert g \vert^2 \vec{d}_{\beta \delta}^\At \cdot \vec{d}_{\alpha \gamma}^{\Dt*}\vec{d}_{\beta \delta}^{\At*} \cdot \vec{d}_{\alpha \gamma}^\Dt\bigr] .
\end{multline}
The projections of donor and acceptor dipole moments can be computed using their spherical components as  defined in the End Matter section, exploiting the fact that each donor dipole matrix elements has only a single spherical vector component, $\vec{d}_{\gamma\alpha}^\Dt =  -d^\Dt_{\pm 1} \ev_{\mp 1}$. One finds \mbox{$\vec{d}_{\beta \delta}^\At \cdot\vec{d}_{\alpha \gamma}^\Dt= d_{\pm 1 \beta \delta}^\At d_{\pm 1}^{\Dt*}$}, \mbox{$\quad\vec{d}_{\beta \delta}^{\At*} \cdot \vec{d}_{\alpha \gamma}^{\Dt*}= d_{\pm 1 \beta \delta}^{\At*}d_{\pm 1}^\Dt$}, \mbox{$\vec{d}_{\alpha \gamma}^{\Dt} \cdot \vec{d}_{\alpha \gamma}^{\Dt*}=|d_{\pm 1}^\Dt|^2$}, \mbox{$\quad\vec{d}_{\beta \delta}^\At \cdot \vec{d}_{\beta \delta}^{\At*}=\sum_{q} d_{q\beta\delta}^\At d_{q\beta\delta}^{\At\ast}$},
\mbox{$\vec{d}_{\beta \delta}^\At\cdot\vec{d}_{\alpha \gamma}^{\Dt*}=-d_{\mp 1 \beta \delta}^\At d_{\pm 1}^\Dt$}, and \mbox{$\vec{d}_{\beta \delta}^{\At*}\cdot\vec{d}_{\alpha \gamma}^\Dt=-d_{\mp 1 \beta \delta}^{\At*}d_{\pm 1}^{\Dt*}$}. As a result, the three projection terms from Eq.~(\ref{eq:11}) can now be expressed explicitly in terms of the donor and acceptor variables:
\begin{eqnarray}
\label{eq:18}
    \sum_{ \psi_i} \vec{d}_{\beta \delta}^\At\cdot\vec{d}_{\alpha \gamma}^\Dt   \vec{d}_{\beta \delta}^{\At*}\cdot \vec{d}_{\alpha \gamma}^{\Dt*}
    &=&\vert d_{\pm 1}^\Dt \vert^2 \frac{\mathrm{d}\sigma^\pm}{\mathrm{d\Omega}},\\
\label{eq:19}
    \sum_{ \psi_i} \vec{d}_{\beta \delta}^\At \cdot \vec{d}_{\beta \delta}^{\At*}\vec{d}_{\alpha \gamma}^\Dt \cdot  \vec{d}_{\alpha \gamma}^{\Dt*}
    &=&\vert d_{\pm 1}^\Dt \vert^2 \sum_q \frac{\mathrm{d}\sigma^q}{\mathrm{d\Omega}},\\
\label{eq:20}
    \sum_{ \psi_i}   \vec{d}_{\beta \delta}^\At\cdot \vec{d}_{\alpha \gamma}^{\Dt*} \vec{d}_{\beta \delta}^{\At*}\cdot \vec{d}_{\alpha \gamma}^\Dt
    &=& \vert d_{\pm 1}^\Dt \vert^2 \frac{\mathrm{d}\sigma^\mp}{\mathrm{d\Omega}}.
\end{eqnarray}
Here, the acceptors' differential photoionization cross section upon absorption of a photon with the polarization $q$ is defined as: $\frac{\mathrm{d}\sigma^q}{\mathrm{d\Omega}}=\sum_{\psi_i} \vert d_{q\beta \delta}^\At \vert^2$. 


\paragraph*{Average over rotational orientations of acceptor.}
As follows from Eqs.~(\ref{eq:18}--\ref{eq:20}), the averaging over relative donor--acceptor orientations needs to be performed only for the differential cross sections  for the ionization of the acceptor molecule, and it straightforwardly yields \cite{Knie14,Ilchen17}:
\begin{equation}
\frac{\mathrm{d}\sigma^\pm}{\mathrm{d\Omega}}=\frac{ {\sigma} }{4\pi}\left[ 1 \pm \beta_1 P_1(\cos \theta)  -\frac{1}{2} \beta_2 P_2(\cos \theta)\right]  \label{eq:DPICSpm}
\end{equation}
for circularly polarized light with $q=\pm1$, and 
\begin{equation}
\frac{\mathrm{d}\sigma^0}{\mathrm{d\Omega}}=\frac{ {\sigma} }{4\pi}\left[ 1 + \beta_2 P_2(\cos \theta)\right] \label{eq:DPICS0}
\end{equation}
for linearly polarized light with $q=0$. Here, $\theta$ is the angle between the direction of the emission of electrons and the direction of the propagation of the ionizing radiation. Explicit analytic expressions for the total photoionization  cross section  $\sigma$, dichroic parameter $\beta_1$ which describes the conventional PECD, and  anisotropy parameter $\beta_2$ can be found in Refs.~\cite{Knie14,Ilchen17}. Equation~(\ref{eq:DPICSpm}) can directly be substituted into Eqs.~(\ref{eq:18}) and (\ref{eq:20}), while combining it with Eq.~(\ref{eq:DPICS0}) reduces the polarization-averaged term in Eq.~(\ref{eq:19}) to the following isotropic result: 
\begin{equation}
\sum_q \frac{\mathrm{d}\sigma^q}{\mathrm{d\Omega}}
=\frac{ {3\sigma} }{4\pi} , \label{eq:CrossTerm}
\end{equation}
which cancels out in the dichroic difference (see below).


\paragraph*{Angle-resolved transition rate.}
Substituting Eqs.~(\ref{eq:18}--\ref{eq:CrossTerm}) into Eq.~(\ref{eq:11}), we arrive at the following angle-resolved transition rate for the antenna-induced ionization
\begin{equation}
\begin{split}
\label{Eq:15x}
     &\Gamma_{\pm}(\theta)={\frac{\vert d_{\pm 1}^\Dt \vert^2}{8\pi\varepsilon_0^2\hbar r^6}} \biggl[\bigl(|f|^2-\tfrac{1}{3}gf^\ast-\tfrac{1}{3}fg^\ast+\tfrac{1}{15}|g|^2\bigr)\\
     &\quad\times\frac{\mathrm{d}\sigma^\pm}{\mathrm{d\Omega}}
     +\tfrac{1}{15}|g|^2\sum_q \frac{\mathrm{d}\sigma^q}{\mathrm{d\Omega}}
     +\tfrac{1}{15}|g|^2\frac{\mathrm{d}\sigma^\mp}{\mathrm{d\Omega}}\biggr]\\
     &=\frac{c^2 \vert d_{\pm 1}^\Dt \vert^2\sigma}{32\pi^2\varepsilon_0^2\hbar\omega_\Dt^4r^6} \biggl\{\bigl(|f|^2-\tfrac{1}{3}gf^\ast-\tfrac{1}{3}fg^\ast+\tfrac{1}{15}|g|^2\bigr)\\
     &\quad\times\bigl[1\pm\beta_1P_1(\cos\theta)-\tfrac{1}{2}\beta_2P_2(\cos\theta)\bigr]+\tfrac{1}{5}|g|^2\\
     &\quad +\tfrac{1}{15}|g|^2\bigl[1\mp\beta_1P_1(\cos\theta)-\tfrac{1}{2}\beta_2P_2(\cos\theta)\bigr]\biggr\}.
\end{split}
\end{equation}
We notice that, for a given helicity of the exciting light as indicated by the subscript $\pm$, the angular distribution of the emitted electrons has three contributions with different weights.  The term $+\tfrac{1}{5}|g|^2$ is angle-independent;  the term $\bigl(|f|^2-\tfrac{1}{3}gf^\ast-\tfrac{1}{3}fg^\ast+\tfrac{1}{15}|g|^2\bigr)$ has the same angular pattern as would result from directly ionizing the chiral molecule with the incoming light (see the `$\pm$' signs in front of the dichroic parameter); and  the term $\tfrac{1}{15}|g|^2$ has the opposite effect, i.e., that would result when directly ionizing the molecule by a photon with the opposite helicity (see the `$\mp$' signs in front of the dichroic parameter). This can be understood from the electrostatic dipole--dipole interaction~(\ref{eq:9}) in a classical picture: In the special case where the intermolecular orientation is parallel to the direction of propagation of the ionizing radiation ($\vec{e}_r=\vec{e_z}$), the donor dipole rotates in the plane perpendicular to the separation ($\hat{\vec{d}}^\Dt\cdot\vec{e}_r=0$) and the dipole--dipole interaction becomes proportional to the product $\hat{\vec{d}}^\At\cdot\hat{\vec{d}}^\Dt$. The acceptor dipole hence has the same rotational sense as if it was directly induced by the circularly polarized light and, thus, produces a PECD of the same sign. This case corresponds to the contribution proportional to $|f|^2$. For all other orientations of the separation vector, the rotating donor dipole induces an acceptor dipole via the dipole--dipole interaction that is given by a coherent superposition of co-rotating, counter-rotating, and non-rotating terms, which, all together, emerge in the final result with different weights upon averaging.

\paragraph*{Dichroic difference in the nonretarded limit.}
In the nonretarded limit, the approximations $f\approx 1$ and $g\approx 3$ can be used to simplify
\begin{multline}
\Gamma_{\pm}(\theta)={\frac{\vert d_{\pm 1 }^\Dt \vert^2\sigma}{32\pi^2\varepsilon_0^2\hbar r^6}}\bigl[2\mp\beta_1P_1(\cos\theta)\\
-\tfrac{1}{10}\beta_2P_2(\cos\theta)\bigr]. \label{eq:FinRate}
\end{multline}
In this limit, the interference term $-\tfrac{1}{3}gf^\ast-\tfrac{1}{3}fg^\ast$ dominates the first term in Eq.~(\ref{Eq:15x}), giving it a negative sign. The net angular distribution of this first term, thus, flips its sign, exhibiting a similar PECD effect as compared to the third term in Eq.~(\ref{Eq:15x}), both opposite to the usual $\pm$ signs  expected for a direct molecular ionization channel. This fact is explicitly indicated by the $\mp$ signs in front of the dichroic parameter $\beta_1$ in Eq.~(\ref{eq:FinRate}). To establish a contact with the photoelectron circular dichroism, we consider  the dichroic difference of the rates for two helicities of light
\begin{equation}
     \Delta \Gamma= \Gamma_+ - \Gamma_- = -{\frac{\vert d_{\pm 1}^\Dt \vert^2\sigma}{16\pi^2\varepsilon_0^2\hbar r^6}}\beta_1P_1(\cos\theta)\label{eq:DichrDif}
\end{equation}
and the respective normalized difference 
\begin{equation}
     \frac{\Gamma_+ - \Gamma_-}{\frac{1}{2}(\Gamma_+ + \Gamma_-)} = -\frac{\beta_1P_1(\cos\theta)}{1-\tfrac{1}{20}\beta_2P_2(\cos\theta)}.\label{eq:NormDif}
\end{equation}
As one can see, the relative strength of the presently uncovered effect, given by the normalized difference (\ref{eq:NormDif}), is in the order of the $\beta_1$ value, which is about two times smaller then the respective relative strength of the conventional PECD effect ($\sim 2\beta_1$). Given that typical values of dichroic parameters emerge on a typical scale of a few to about ten percent, one can expect a similarly-sizable relative strength for the presently predicted antenna-induced PECD (see below).


\paragraph*{Possible experiment and feasibility.}
While most rICD  studies so far have considered homogeneous or heterogeneous rare-gas dimers or clusters \cite{Trinter2013,Hans2019,Barth2005}, this field was recently extended to dimers of two organic molecules \cite{Bejoy2023}. It is, thus, worth investigating the photoelectron angular distribution of  molecular complexes, and due to their simpler and known electronic structure, particularly atom--molecule complexes across atomic resonances, which are  energetically embedded in the ionization continuum of chiral molecules. Rare gas atoms are  well suited for this, since they exhibit a variety of resonances in the ionization continuum of typical organic (chiral) molecules. Such complexes of rare gas atoms and organic molecules have been recently experimentally produced and investigated \cite{Bohlen2022}. A simple model system could, e.g., be a complex of a He atom and a camphor molecule, experimentally realized as camphor-doped He nanodroplets \cite{Sen2024}. One can tune the circularly polarized light to the He$^\ast(1s^12p^1~^1P)$ excitation, selectively populating the $m=\pm1$ states of the $2p$ electron  with different light helicities. The respective resonance energy of about 21.218~eV \cite{Nist2024} is well above the ionization energy of camphor (8.7\,eV \cite{Rennie2002}).  At this photon energy, the dichroic parameter for the direct ionization of the \textit{R-}enantiomer of camphor is equal to about $\beta_1 =-6\%$ \cite{Powis2008b,Hadidi18}, and the conventional  ${\mathrm{PECD} =~} 2\beta_1 =-12\%$, with fewer photoelectrons emitted in the forward and more in the backward direction with respect to the propagation direction of the circularly polarized ionizing light of positive helicity. According to Eq.~(\ref{eq:NormDif}), one expects a half as large asymmetry  with opposite sign  ${\mathrm{PECD} =~}-\beta_1 =+6\%$ for the RET-induced ionization of \textit{R-}camphor, with slightly more electrons emitted in the forward then in the backwards directions. A differentiation of the rICD-induced PECD effect for molecular complexes of various sizes, simultaneously present in a typical target, is an experimental challenge which can possibly be solved involving coincidence detection methods. Such mass-selected PECD measurements, i.e., photoelectron angular distributions in coincidence with specific molecular complexes, have recently been demonstrated \cite{Powis2014,Rouquet2023}. 


\paragraph*{Summary.} 
We have shown that the excitation of an achiral antenna atom with circularly polarized light can induce a PECD-like signal in a nearby chiral molecule by means of rICD. Hereby, the information of the rotatory sense of the circularly polarized field is transmitted to the molecule via a (retarded) dipole--dipole interaction. We have derived an effective photoelectron angular distribution that results from independent averages of the orientations of the intermolecular line-of-sight and the donor molecule. With respect to the former, the rotating dipole of the atom induces  molecular dipoles that have the same or opposite senses of rotation with distance-dependent weights. In the near-field limit, the antenna-induced PECD has the opposite sign as compared to that if an incoming photon would directly ionize the molecule. In view of possible coincidence experiments, photoelectron distributions for given intermolecular line-of-sight may be an interesting subject for further investigations. More fundamentally, interference effects between the dominant antenna-induced and weak direct PECDs \cite{Hartmann2019} would be worth considering. The exemplary  rICD phenomenon investigated here suggests a range of further antenna-induced photoionization processes, e.g.,  where both constituents are chiral.


\begin{acknowledgments}
S.Y.B. would like to thank Christiane Koch for valuable discussions. This work was funded by the Deutsche Forschungsgemeinschaft (DFG, German Research Foundation) -- Project No. 328961117 -- SFB 1319 ELCH (Extreme light for sensing and driving molecular chirality). 
\end{acknowledgments}



\begin{widetext}
\begin{center}
\large{\textbf{End Matter}}
\end{center}
\end{widetext}

\textit{Macroscopic QED}---Macroscopic QED is a theory of the interaction of individual particles with charges $q_i$ with the quantized electromagnetic field in the possible presence of macroscopic bodies or media. When grouping the particles into neutral molecular systems with center-of-mass positions $\mathbf{r}_M$, the interaction of each system with the field can be described by a multipolar Hamiltonian \cite{Buhmann04}
\begin{equation} \tag{EM.1}
\label{EqA1}
    \hat{H}_{M\mathrm{F}}=-\int\mathrm{d}^3r\,\hat{\vec{P}}_M(\vec{r})\sprod\hat{\vec{E}}(\vec{r}),
\end{equation}
where
\begin{multline}\tag{EM.2}
\label{EqA2}
    \hat{\vec{P}}_M(\vec{r})=\sum_{i\in M}q_i(\hat{\vec{r}}_i-\vec{r}_M)\\
    \times\int_0^1\mathrm{d}\lambda\,\delta[\vec{r}-\vec{r}_M-\lambda(\hat{\vec{r}}_i-\vec{r}_M)]
\end{multline}
is the molecular polarization and we have neglected magnetic and R\"ontgen interactions. In contrast to the equivalent minimal-coupling description, molecular systems interact only via their coupling to the field. In particular, the electrostatic Coulomb interaction
\begin{equation} \tag{EM.3}
\label{EqA2x}
    \hat{H}_{MM'}=\sum_{i\in M}\sum_{j\in M'}\frac{q_iq_j}{4\pi\varepsilon_0|\hat{\vec{r}}_i-\hat{\vec{r}}_j|}
\end{equation}
of charges across different molecular systems is implicitly included as part of the full retarded coupling $\hat{H}_{M\mathrm{F}}+\hat{H}_{M'\mathrm{F}}$ \cite{Buhmann04}. Assuming that the size of each molecular system is much smaller than the wavelengths of the relevant photons emitted and absorbed, the multipolar Hamiltonian reduces to its electric dipole form in leading-order long-wavelength approximation
\begin{equation} \tag{EM.4}
\label{EqA3}
    \hat{H}_{M\mathrm{F}}=-\hat{\vec{d}}_M\sprod\hat{\vec{E}}(\vec{r}_M)
\end{equation}
with $\hat{\vec{d}}_M=\sum_{i\in M}q_i(\hat{\vec{r}}_i-\vec{r}_M)$ being the electric dipole moment of the respective molecular system.

The electric-field operator in the possible presence of dispersing and absorbing media with dielectric permittivity $\varepsilon(\vec{r},\omega)$ can be given as
\begin{multline} \tag{EM.5}
\label{EqA4}
    \hat{\vec{E}}(\vec{r})=\mathrm{i}\sqrt{\frac{\hbar}{\pi\varepsilon_0}}\int_0^\infty\mathrm{d}\omega \frac{\omega^2}{c^2}
\int\mathrm{d}^3r'\sqrt{\operatorname{Im}\varepsilon(\vec{r}',\omega)}\\
    \times\tens{G}(\vec{r},\vec{r}',\omega)\sprod\hat{\vec{f}}(\vec{r}',\omega)+\operatorname{h.c.}
\end{multline}
Here, the Green tensor is the solution to a Helmholtz equation:
\begin{equation}\tag{EM.6}
\label{EqA5}
\biggl[\bm{\nabla}\vprod\bm{\nabla}\vprod\,-\frac{\omega^2}{c^2}\varepsilon(\vec{r},\omega)\biggr]\tens{G}(\vec{r},\vec{r}',\omega)=\bm{\delta}(\vec{r}-\vec{r}').
\end{equation}
It serves as a propagator for the electric field. The bosonic creation and annihilation operators $\hat{\vec{f}}^\dagger(\vec{r},\omega)$ and $\hat{\vec{f}}(\vec{r},\omega)$ represent the fluctuating noise polarization within the media. 

As seen from the Helmholtz equation~(\ref{EqA5}), the full interaction~(\ref{EqA3}) accounts for both retardation and environments via the Green tensor. It hence operates at relatively large distances. In particular, the multipolar coupling scheme is typically based on the assumption
\begin{equation}\tag{EM.7}
\label{EqA6}
    \int\mathrm{d}^3r\,\hat{\vec{P}}_M(\vec{r})\sprod\hat{\vec{P}}_{M'}(\vec{r})=0
\quad\mathrm{for}\,\,M\neq M',
\end{equation}
hence neglecting the effect of electronic wave-function overlap across molecules. On the opposite end of the intermolecular distance range, the molecule--molecule interaction in free space in the nonretarded, electrostatic limit reduces to the Coulomb interaction~(\ref{EqA2x}), which does account for molecular wave-function overlap and hence also charge transfer processes and exchange interactions. A hierarchy of interaction Hamiltonians for ascending distances can hence be given as $\hat{H}_{MM'}$ -- $\hat{V}$ -- $\hat{H}_{M\mathrm{F}}$. The electrostatic dipole--dipole interaction $\hat{V}$, as given by Eq.~(\ref{eq:9}) in the main text, can be approached  from $\hat{H}_{MM'}$ as a leading-order multipole expansion in the limit of distances much larger than the size of the molecular systems, or from $\hat{H}_{M\mathrm{F}}$ in the limit of distances much smaller than the wavelengths of photons, associated with the relevant molecular transitions. Our calculation is based on $\hat{H}_{M\mathrm{F}}$, where the special case $\hat{V}$ is included for illustrative purposes. 

\textit{Resonance energy transfer}---The fully retarded resonance energy transfer rate~(\ref{eq:1}) can be obtained from Fermi's golden rule in the form
\begin{equation}\tag{EM.8}
\label{EqA7}
   \Gamma_{i}=\frac{2\pi}{\hbar}\sum_{f}|M_{fi}|^2\delta(E_i-E_f)
\end{equation}
with initial state $\ket{i}=\ket{\gamma}\ket{\delta}\ket{\{0\}}$ and final state $\ket{f}=\ket{\alpha}\ket{\beta}\ket{\{0\}}$. Here, $\ket{\{0\}}$ denotes the vacuum state of the electromagnetic field as defined by $\hat{\vec{f}}(\vec{r},\omega)\ket{\{0\}}=0$. To the leading (second order) in the molecule--field coupling $\hat{H}=\sum_{M=\mathrm{D},\mathrm{A}}\hat{H}_{M\mathrm{F}}$ (where D\,=\,donor and A\,=\,acceptor), the transition matrix element is given by
\begin{equation}\tag{EM.9}
    M_{fi}=\lim_{\epsilon\to 0_+}\sum_m\frac{\bra{f}\hat{H}\ket{m}\bra{m}\hat{H}\ket{i}}{E_i-E_m-\mathrm{i\epsilon}}\,,
\end{equation}
where the two possible intermediate states $\ket{m}=\ket{\alpha}\ket{\delta}\ket{\bm{1}(\vec{r},\omega},\ket{\beta}\ket{\gamma}\ket{\bm{1}(\vec{r},\omega}$ include single-photon excitations $\ket{\bm{1}(\vec{r},\omega)}=\hat{\vec{f}}^\dagger(\vec{r},\omega)\ket{\{0\}}$. The matrix element can be evaluated by using the electric-dipole coupling~(\ref{EqA3}) with the electric field~(\ref{EqA4}). As shown in Ref.~\cite{Hemmerich2018}, one finds
\begin{equation}\tag{EM.10}
    M_{fi}=-\frac{\omega_\mathrm{D}^2}{\varepsilon_0c^2}\vec{d}_{\beta \delta}^\At\sprod \tens{G} (\vec{r}_\At,\vec{r}_\Dt,\omega_\Dt) \sprod\vec{d}_{\alpha \gamma}^\Dt
\end{equation}
for $\omega_\mathrm{A}=\omega_\mathrm{D}$. Substitution of this result into Fermi's golden rule~(\ref{EqA7}) leads to the result~(\ref{eq:1}) of the main text.

\textit{Spherical vectors}---Introducing spherical unit vectors $\ev_0 = \ev_z$, $\ev_{\pm1}=\mp (\ev_x \pm \text{i}\ev_y)/\sqrt{2}$ with orthonormality $\ev_q \cdot \ev_{q'}= (-1)^q \delta_{q, -q'}$, one can expand arbitrary vectors $\vec{v}= \sum_{q}(-1)^q v_{-q} \ev_q$ into their associated spherical vector components $v_0 = v_z$, $v_{\pm1}=\mp (v_x \pm \text{i}v_y)\sqrt{2}$. Thereby, a complex conjugation of unit vectors and components obeys the identities $\ev_q^* = (-1)^q \ev_{-q}$, $v_q^* = (-1)^q v_{-q}$, and the scalar product of two vectors $\vec{v}$ and $\vec{w}$ can be expressed in terms of spherical components according to $\vec{v}\cdot \vec{w}= \sum_{q}(-1)^q v_{-q} w_q$.

\end{document}